\documentclass{article}

\usepackage{latexsym,amsmath,amssymb,amscd,amsfonts,amsthm,enumerate}

\begin{document}

\title{Similarity solutions and conservation laws for the Beam Equations: a complete study}

\author{Amlan K Halder\thanks{%
Email: amlan91.res@pondiuni.edu.in} \\
{\ \ \textit{Department of Mathematics, Pondicherry University,}}\\
{\ \textit{Kalapet, India-605014}}
\and Andronikos Paliathanasis\thanks{%
Email: anpaliat@phys.uoa.gr} \\
{\ \textit{Instituto de Ciencias F\'{\i}sicas y Matem\'{a}ticas, }}\\
{\textit{Universidad Austral de Chile, Valdivia, Chile and}}\\
{\textit{Institute for Systems Science, Durban University of Technology}}\\
{\textit{Durban 4000, Republic of South Africa.}}\\
\and PGL Leach\thanks{%
Email: leachp@ukzn.ac.za} \\
{\textit{School of Mathematics, Statistics and Computer Science,}}\\
{\textit{University of KwaZulu-Natal, Durban, South Africa and}}\\
{\textit{Institute for Systems Science, Durban University of Technology}}\\
{\textit{Durban 4000, Republic of South Africa.}}}

\maketitle

\begin{abstract}
We study the similarity solutions and we determine the conservation laws of the various forms of beam equation, such as, Euler-Bernoulli, Rayleigh and Timoshenko-Prescott. The travelling-wave reduction leads to solvable fourth-order odes for all the forms. In addition, the reduction based on the scaling symmetry  for the Euler-Bernoulli form leads to certain odes for which there exists zero symmetries. Therefore, we conduct the singularity analysis to ascertain the integrability. We study two reduced odes of order second and third. The reduced second-order ode is  a perturbed form of Painlev\'e-Ince equation, which is integrable and the third-order ode falls into the category of equations studied by Chazy, Bureau and Cosgrove. Moreover, we derived the symmetries and its corresponding reductions and conservation laws for the forced form of the above mentioned beam forms. The Lie Algebra is mentioned explicitly for all the cases.

[MSC 2010]{34A05; 34A34; 34C14; 22E60; 35B06; 35C05; 35C07}

\end{abstract}

\vspace{1.5cc}

\section{Introduction}

There are basically two type of beam. One type is supported at both ends and
the other type is supported at one end only. It is a cantilever. The latter
is of greater mathematical and physical interest for the free end can
vibrate. This causes stresses in the beam. The first mathematical
description was made by Leonhard Euler and Daniel Bernoulli around $1750$, but
there were some earlier attempts by Leonardo da Vinci and Galileo Galilei
who were more than a little hampered by no knowledge of differential
equations. Jacob Bernoulli laid the groundwork for the development of
Leonhard Euler and Daniel Bernoulli. In $1894$, the polymath Lord Rayleigh
proposed an improvement to the Euler-Bernoulli model by including a term
related to rotational stress. In $1921$, Timoshenko introduced considerable
improvements in what is now termed the Timoshenko-Prescott model.

There has been considerable experimental and numerical work devoted to the
comparison of predictions of the theories and experimental results. It
should be emphasised that the infinitesimal theory of elasticity is
three-dimensional and that the three models mentioned above are linear
models. They make for easier mathematics, but there is a price to pay.
Curiously the simplest model, that of Euler-Bernoulli, still finds favour
amongst some practitioners.

Some of the experimental work \cite{Davies 48 a} undertaken to compare
reality with theoretical prediction tries to make the experiment as close to
a one-dimensional model as is possible. One of the more interesting studies
is the propagation of shock waves along the beam. This involves firing a
bullet into the fixed end of the cantilever which is of slight diameter --
25 mm -- to emulate a uniform boundary condition at the fixed end of the
beam.

The literature devoted to the theory and practice of beams is extensive both
in time and space. A fairly recent paper by Labuschagne \cite%
{Labuschagne 09 a} is very good in its historical aspects as well as being
clearly written. Earlier papers in addition to that of Davies cited above
are by Hudson \cite{Hudson 43 a} and Bancroft \cite{Bancroft 41 a}. An
interesting feature is that the beams are taken to be cylindrical in shape
even though the beams one sees in buildings are anything but cylindrical
with some exceptions to be found in beamed structures of the nineteenth
century. One assumes that this makes the analysis simpler due to the radial
symmetry. Even a square cross section would complicate the mathematics
seriously.\\

In this work we study the algebraic properties of the Euler-Bernoulli, the
Rayleigh and of the Timoshenko-Prescott according to the admitted Lie point
symmetries, for the source-free equation as also in the case where a
homogeneous source term exists. The application of symmetry analysis for the
Euler-Bernoulli equation is not new, there are various studies in the
literature \cite{b4,b1,b2,b5,b3}, however in this paper, we obtained some
new results, as the reduction of Euler-Bernoulli form to perturbed form of
Painlev\'e-Ince \cite{Ince 27 a} equation, which is integrable and the third-order ode
which falls into the category of equations studied by Chazy, Bureau and
Cosgrove. Also, we show that the three beam equations of our study
admit the same travelling-wave solution.\\

Certain phenomenal works were recently done for the static Euler-Bernoulli
equation by Ruiz \cite{Acta03} and  Da Silva PL \cite{Acta02}. In Ruiz \cite{Acta03} the Euler-Bernoulli equation with an external agent is studied
with respect to the joint invariants of the algebra and complete solutions are specified whereas in \cite{Acta02} for the static Euler-Bernoulli equation with specific nonlinear term it was found that the algebraic structure of Lie point symmetries is similar to that of the Noether symmetries. It is worthwhile to mention the paper of Freire IL \cite{Acta01} where the Lane-Emden system is reduced to the Emden-Fowler equations and correspondingly the solutions of the system have been studied with the aid of the point symmetries. To elaborate the above mentioned works we focus on the more general Euler-Bernoulli equation with and without the external forcing term and compute its solution using the point symmetries. It is also our intuition that the point symmetries of the form of Euler-Bernoulli under consideration do possess some similarities with the Noether symmetries to follow the results of the above mentioned work.\\

This paper is structured as: In Section $2$, we mention the Lie point
symmetries and the corresponding algebra. In Section $3$, we discuss the
travelling-wave solutions for all the beam forms and further reductions of
Euler-Bernoulli form using the scaling symmetries. In Section $4$, we study
the forced forms of the beam equations. Section 5 is devoted to the
singularity analysis of a third-order equation, which is obtained by the
reduction of Euler-Bernoulli equation by using the scaling symmetry.
Conservation laws for the three beam equations are derived in Section $6$. The
conclusion and appropriate references are mentioned henceforth.

\section{Lie symmetry analysis}

For the convenience of the reader, we give a briefly discussion in the
theory of Lie point symmetries. In particular, we present the basic
definitions and main steps for the determination of Lie point symmetries for
a give differential equation. Consider $H^{A}\left( t,x,u,u_{,i}\right) =0$
to be a set of differential equations$~$and $u_{,i}=\frac{\partial u}{%
\partial y^{i}}$ in which $y^{i}=\left( t,x\right) $.

Then under the action of the infinitesimal one-parameter point
transformation
\begin{eqnarray}
t^{\prime } &=&t\left( t,x,u;\varepsilon \right),  \label{ls.01} \\
x^{\prime } &=&x\left( t,x,u;\varepsilon \right),  \label{ls.02} \\
u^{^{\prime }A} &=&u^{A}\left( t,x,u;\varepsilon \right),  \label{ls.03}
\end{eqnarray}%
in which $\varepsilon $ is an infinitesimal parameter, the set of
differential equations $H^{A}$ is invariant if and only if,
\begin{equation}
H^{A}\left( t^{\prime },x^{\prime },u^{\prime }\right) =H^{A}\left( t,x,u\right),
\label{ls.04}
\end{equation}%
or equivalently%
\begin{equation}
\lim_{\varepsilon \rightarrow 0}\frac{H^{A}\left( t^{\prime },x^{\prime
},u^{\prime };\varepsilon \right) -H^{A}\left( t,x,u\right) }{\varepsilon }%
=0.  \label{ls.05}
\end{equation}%
The later expression is the definition of the Lie derivative $\mathcal{L}~$%
of $H^{A}$ along the direction
\begin{equation}
\Gamma =\frac{\partial t^{\prime }}{\partial \varepsilon }\partial _{t}+%
\frac{\partial x^{\prime }}{\partial \varepsilon }\partial _{x}+\frac{%
\partial u^{\prime}}{\partial \varepsilon }\partial _{u}.  \label{ls.06}
\end{equation}

Hence, we shall say that the vector field $\Gamma $ will be a Lie point
symmetry for the set of differential equations $H^{A}$ if and only if the
following condition is true%
\begin{equation}
\mathcal{L}_{\Gamma }\left( H^{A}\right) =0.  \label{ls.07}\\
\end{equation}

In other words, the operator $\Gamma$ can be considered to be  symmetry provided
\begin{equation*}
\Gamma^{[n]} H^{A} =0,
\end{equation*}
whenever $H^{A}(t,x,u,u,_{i})=0$ and $\Gamma^{[n]}$ denotes the $n$-th prolongation of the specified operator in its defined space. The set of all such operators can be denoted by {\bf {G}} which can be regarded as the symmetry group for the set of differential equations  $H^{A}(t,x,u,u,_{i})=0$ \cite{U02,U03,U01}.\\

\subsection{The Euler-Bernoulli equation.\newline}

The Euler-Bernoulli form of the beam equation is \cite{Davies 48 a,beam01},
\begin{equation}
\alpha \beta u_{xxxx}+u_{tt}=0.  \label{01a}
\end{equation}%
The Lie point symmetries are
\begin{eqnarray}
\Gamma _{1a} &=&\partial _{x}~,~\Gamma _{2a}=\partial _{t},  \notag \\
\Gamma _{3a} &=&u\partial _{u}~,~\Gamma _{4a}=2t\partial _{t}+x\partial _{x},
\notag \\
\Gamma _{5a} &=&a(t,x)\partial _{u},  \notag
\end{eqnarray}%
where $a(t,x)$ satisfies the Euler-Bernoulli form of beam equation. The Lie
Algebra is $(A_{3,3}\oplus A_{1})\oplus _{s}\infty A_{1}$, according to the
Morozov-Mubarakzyanov classificatin scheme \cite{burg 21, burg 22, burg 23,
burg 24}. \footnote{We use the SYM package developed by Prof.Stelios Dimas \cite{Dimas 04 a,Dimas 06}.}

\subsection{The Rayleigh equation.\newline}

The Rayleigh form of the beam equation is \cite{Davies 48 a,beam01},%
\begin{equation}
\alpha \beta u_{xxxx}+u_{tt}-\beta u_{xxtt}=0.  \label{01c}
\end{equation}%
The Lie point symmetries are
\begin{eqnarray}
\Gamma _{1b} &=&\partial _{t}~,~\Gamma _{2b}=\partial _{x},  \notag \\
\Gamma _{3b} &=&u\partial _{u}~,~\Gamma _{4b}=b(t,x)\partial _{u},  \notag
\end{eqnarray}%
where $b(t,x)$ satisfies the Rayleigh form of the beam equation.
Consequently, the admitted Lie algebra is $A_{3}\oplus _{s}\infty A_{1}$.%
\newline

\subsection{The Timoshenko-Prescott equation.\newline}

The Timoshenko and Prescott form of the beam equation is \cite{Davies 48 a,beam02},%
\begin{equation}
\alpha \beta u_{xxxx}+u_{tt}-\beta (1+\epsilon )u_{xxtt}+\frac{\epsilon
\beta u_{tttt}}{\alpha }=0.  \label{01e}
\end{equation}%
The Lie point symmetries are
\begin{eqnarray}
\Gamma _{1c} &=&\partial _{t}~,~\Gamma _{2c}=\partial _{x}~,~\Gamma
_{3c}=u\partial _{u},  \notag \\
\Gamma _{4c} &=&c(t,x)\partial _{u},  \notag
\end{eqnarray}%
where $c(t,x)$ satisfies the Timoshenko and Prescott form of the beam
equation. Hence, the admitted Lie Algebra is $A_{3}\oplus _{s}\infty A_{1}$.

Therefore, we can say that the Timoshenko-Prescott equation and the Rayleigh
equation are algebraic equivalent, but different with the Euler-Bernoulli
equation admit a higher-dimensional Lie algebra.

We proceed our analysis by applying the Lie point symmetries to determine
similarity solutions for the three equations of our study.

\section{The travelling-wave solution}

The travelling-wave solution for eq (\ref{01a}), with respect to $\Gamma
_{2a}+c\Gamma _{1a}$, where $c$, denotes the frequency, leads to the fourth
order equation,
\begin{equation}
c^{2}v^{\prime \prime }(s)+\alpha \beta v^{\prime \prime \prime \prime
}(s)=0,  \label{02a}
\end{equation}%
where
\begin{eqnarray*}
s &=&x-ct, \\
v(s) &=&u(x,t).
\end{eqnarray*}%
The Lie point symmetries of equation (\ref{02a}) are
\begin{eqnarray}
\Gamma _{1d} &=&\partial _{s},  \notag  \label{02b} \\
\Gamma _{2d} &=&\partial _{v},  \notag \\
\Gamma _{3d} &=&s\partial _{v},  \notag \\
\Gamma _{4d} &=&v\partial _{v},  \notag \\
\Gamma _{5d} &=&\sin {\frac{cs}{\sqrt{\alpha \beta }}}\partial _{v},  \notag
\\
\Gamma _{6d} &=&\cos {\frac{cs}{\sqrt{\alpha \beta }}}\partial _{v}.  \notag
\end{eqnarray}

The reduced form has six symmetries and hence linearisable, the solution for
the fourth-order equation can be given as,
\begin{equation}
v(s)=C_{0}+C_{1}s+C_{2}\sin {\frac{cs}{\sqrt{\alpha \beta }}}+C_{3}\cos {%
\frac{cs}{\sqrt{\alpha \beta }}},  \notag  \label{02c}
\end{equation}%
where $C_{i}$, $i=0,1,2,3$ are arbitrary constants. Corresponding, to which
the solution for Euler-Bernoulli form of beam can be given as
\begin{equation}
u(x,t)=C_{0}+C_{1}(x-ct)+C_{2}\sin {\frac{c(x-ct)}{\sqrt{\alpha \beta }}}%
+C_{3}\cos {\frac{c(x-ct)}{\sqrt{\alpha \beta }}}.  \notag  \label{02c1}
\end{equation}

\subsection{Further reduction of the Euler-Bernoulli equation.\newline
}

The reduction with respect to $\Gamma _{3a}$ and $\Gamma _{4a}$, leads to
fourth-order odes. For the similarity variables with respect to $2\Gamma _{3a}$+$\Gamma _{4a}$,
\begin{eqnarray}
s &=&\frac{t}{x^{2}},  \notag  \label{02d} \\
u(t,x) &=&tv(s),  \notag
\end{eqnarray}%
the reduced ode is
\begin{equation}
\left( \frac{2}{\alpha \beta }+120s^{2}\right) v^{\prime }+\left(
s+300s^{3}\right) v^{\prime \prime }+144s^{4}v^{\prime \prime \prime
}+16s^{5}v^{\prime \prime \prime \prime }=0.  \notag  \label{02e}
\end{equation}%
The latter equation is solvable. We continue by considering the similarity
variable $u(t,x)=xv(s)$, with respect to $\Gamma _{3a}$+$\Gamma _{4a}$ for $s$ being same as above leads to the
fourth-order ode,
\begin{equation}
24sv^{\prime }+\left( \frac{1}{\alpha \beta }+156s^{2}\right) v^{\prime
\prime }+16s^{3}\left( 7v^{\prime \prime \prime }+sv^{\prime \prime \prime
\prime }\right) =0.  \label{02f}
\end{equation}%
The equation has a total of five Lie point symmetries with $\partial _{v}$, $%
v\partial _{v}$ are the two simpler symmetries, the other three are in terms
of Hypergeometric functions, which is complicated enough to be mentioned
here.

We apply $\partial _{v}$ to perform the reduction. The new invariant
functions are $s=h$ and $g(h)=v^{\prime }(s)$, hence, the reduced equation
is a third-order ode,
\begin{equation}
g^{\prime \prime \prime }(h)=-\frac{7g^{\prime \prime }(h)}{h}-\left( \frac{%
39}{4h^{2}}+\frac{1}{16h^{4}\alpha \beta }\right) g^{\prime }(h)-\frac{3g(h)%
}{2h^{3}}.  \label{02g}
\end{equation}

The latter equation admit four Lie point symmetries. The simpler one being $%
g\partial _{g}$, the other three are hyperbolic functions of $\sin {h}$,$%
\cos {h}$ and Hypergeometric function respectively. We consider $g\partial
_{g}$, to do the reduction.

The subsequent second-order equation is,
\begin{equation}
m^{\prime \prime }(n)=\left(-3m(n)-\frac{7}{n}\right)m^{\prime}(n)-m(n)^{3}-\frac{7m(n)^{2}}{n}%
-\left( \frac{39}{4n^{2}}+\frac{1}{16n^{4}\alpha \beta }\right) m(n)-\frac{3%
}{2n^{3}},  \label{02h}
\end{equation}%
where $n=h$ and $m(n)=\frac{g^{\prime }(h)}{g(h)}$. This is the perturbed
form of Painlev\'e-Ince equation and the singularity analysis of this
equation shows that it is integrable. In our subsequent paper we look at the
analysis and discuss it elaborately. The reduction of (\ref{02f}) with
respect to $v\partial _{v}$, leads to a third-order equation with zero
symmetries,
\begin{eqnarray}
g^{\prime \prime \prime }(h) &=&\left(-4g(h)-\frac{7}{h}\right)g^{\prime \prime
}(h)-3g^{\prime 2}-\left( 6g(h)^{2}+21\frac{g(h)}{h}+\frac{39}{4h^{2}}+\frac{%
1}{16h^{4}\alpha \beta }\right) g^{\prime }(h)\notag   \\
&&-g(h)^{4}-\frac{7g(h)^{3}}{h}-\left( \frac{39}{4h^{2}}+\frac{1}{%
16h^{4}\alpha \beta }\right) g(h)^{2}-\frac{3g(h)}{2h^{3}}, \label{02i}
\end{eqnarray}%
where $h=s$ and $g(h)=\frac{v^{\prime }(s)}{v(s)}$. This equation is
integrable as ascertained by singularity analysis. The calculations of which
are mentioned in a following section.

\subsection{The Travelling wave solution for the Rayleigh equation.\newline
}

The reduction using $\Gamma _{1b}+c\Gamma _{2b}$ leads to the fourth-order
equation which is maximally symmetric, where $c$ is the frequency. The
definition of similarity variables $s$ and $v(s)$ is similar to that of the
previous case.
\begin{equation}
\left( 1-\beta c^{2}\right) v^{\prime \prime \prime \prime
}(s)+c^{2}v^{\prime \prime }(s)=0,  \label{02j}
\end{equation}%
which is in the form of equation (\ref{02a}).

\subsection{The travelling-wave solution for the Timoshenko-Prescott
equation.\newline
}

In a similar way, the application of the generic symmetry vector, $\Gamma
_{1c}+c\Gamma _{2c}$, in (\ref{01e}) provide the fourth-order ode,
\begin{equation}
\left( \alpha ^{2}\beta -\beta c^{2}a-\alpha \beta c^{2}\varepsilon
+c^{4}\varepsilon \beta \right) v^{\prime \prime \prime \prime
}(s)+ac^{2}v^{\prime \prime }\left( s\right) =0,  \label{02m}
\end{equation}%
which is again in the form of (\ref{02a}). Consequently, we conclude that the
three different beam equations provide the same travel-wave solutions.

We continue our analysis by assuming the existence of a source term $f\left(
u\right) $ in the beam equations.

\section{Symmetry analysis with a source term}

In this section, we study the impact of the forcing-source term $f(u)$ in
the rhs of the Euler-Bernoulli, Rayleigh and Timoshenko-Prescott beam
equations.\newline

\subsection{Euler-Bernoulli.}

The Lie symmetry analysis for the Euler-Bernoulli equation (\ref{01a}) with
forced term $f(u)$, leads to the following possible cases for the forcing
term$~$%
\begin{equation}
f_{1}\left( u\right) =au+b,
\end{equation}%
\begin{equation}
f_{2}\left( u\right) =(au+b)^{n},
\end{equation}%
\begin{equation}
f_{3}\left( u\right) =e^{au+b},
\end{equation}%
\begin{equation}
f_{4}\left( u\right) =\text{arbitrary.}
\end{equation}

For $f_{1}(u)$, the admitted Lie point symmetries for the Euler-Bernoulli
equation are,%
\begin{equation*}
\Gamma _{1}^{f_{1}}=\partial _{t}~,~\Gamma _{2}^{f_{1}}=\partial
_{x}~,~\Gamma _{3}^{f_{1}}=u\partial _{u}~,~\Gamma _{\infty
}^{f_{1}}=b\left( t,x\right) \partial _{u},
\end{equation*}%
where they form the $3A_{1}$ Lie Algebra and $b\left( t,x\right) $ is a
solution of the original equation.

For the source $f_{2}\left( u\right), $ the admitted Lie point symmetries are,%
\begin{equation*}
\Gamma _{1}^{f_{2}}=\partial _{t}~,~\Gamma _{2}^{f_{2}}=\partial
_{x}~,~\Gamma _{3}^{f_{2}}=2\left( n-1\right) t\partial _{t}+\left(
n-1\right) x\partial _{x}-4\left( u+\frac{b}{a}\right) \partial _{u},~
\end{equation*}%
which they form the $2A_{1}\oplus _{s}A_{1}$ Lie Algebra.

For $f_{3}\left( u\right) $ the admitted Lie point symmetries are,%
\begin{equation*}
\Gamma _{1}^{f_{3}}=\partial _{t}~,~\Gamma _{2}^{f_{3}}=\partial
_{x}~,~\Gamma _{3}^{f_{3}}=2t\partial _{t}+x\partial _{x}-\frac{4}{a}%
\partial _{u},
\end{equation*}%
where the corresponding Lie Algebra is the\ $2A_{1}\oplus _{s}A_{1}.$

Finally for arbitrary functional form of $f\left( u\right) $ the admitted
Lie point symmetries are  the only two symmetry vectors,
\begin{equation*}
\Gamma _{1}^{f_{4}}=\partial _{t}~,~\Gamma _{2}^{f_{4}}=\partial _{x},
\end{equation*}%
which form the $2A_{1}~$Lie Algebra and provide the travelling-wave solution.

\subsection{Rayleigh and Timoshenko-Prescott equations.\newline}

For the other two beam equations, namely the Rayleigh and
Timoshenko-Prescott equations with \ a source term, we find that for a linear
function $f=f_{1}\left( u\right) $, the two equations admit the same Lie
point symmetries with the force-free case, while for arbitrary function $%
f\left( u\right) =f_{4}\left( u\right) $, admit only two Lie point
symmetries, the vector fields $\Gamma _{1}^{f_{4}},~\Gamma _{2}^{f_{4}}$
which provide with travelling-wave solutions.

\subsection{Symmetry classification of ODE.\newline}

As we show above the reduction with the Lie point symmetries, $\Gamma
_{1}^{f_{4}}+c\Gamma _{2}^{f_{4}}$, for the three beam equations of our
consideration provide the same fourth-order ODE, which now with a source
term it takes the following form,%
\begin{equation}
v^{\prime \prime \prime \prime }+c^{2}v^{\prime \prime }=f\left( v\right),
\label{ll.01}
\end{equation}

We perform the symmetry classification of the latter differential equation
and we find that for arbitrary function $f\left( v\right) $, the latter
equation admits only the autonomous symmetry vector $\partial _{v}$. However,
for a constant source $f\left( v\right) =a_{0},$ the Lie point symmetries are,%
\begin{equation*}
\partial _{s}~,\partial _{v}~,~s\partial _{v}~,~\left(
2c^{2}v-a_{0}s^{2}\right) \partial _{v}~,~\cos \left( cs\right) \partial
_{v}~,~\sin \left( cs\right) \partial _{v},
\end{equation*}%
where the generic solution of equation (\ref{ll.01}) is,
\begin{equation}
v\left( s\right) =v_{1}\sin \left( cs\right) +v_{2}\cos \left( cs\right)
+v_{3}s+v_{4}+\frac{a_{0}}{2c^{2}}s^{2}.
\end{equation}

On the other hand, for $f\left( v\right) =a_{1}v+a_{0}$, equation~(\ref%
{ll.01}) admits the six Lie point symmetries,%
\begin{equation*}
\partial _{s}~,~\left( a_{1}v+a_{0}\right) \partial _{v}~,~\exp \left( \pm i%
\frac{\sqrt{2c^{2}+2\sqrt{c^{4}+4a_{1}}}}{2}s\right) \partial _{v}~,~\exp
\left( \pm i\frac{\sqrt{2c^{2}-2\sqrt{c^{4}+4a_{1}}}}{2}s\right)\partial_{v},
\end{equation*}%
where the generic solution of (\ref{ll.01})  is,
\begin{eqnarray}
v\left( s\right) &=&v_{1}~\exp \left( i\frac{\sqrt{2c^{2}+2\sqrt{c^{4}+4a_{1}%
}}}{2}s\right) +v_{2}\exp \left( -i\frac{\sqrt{2c^{2}+2\sqrt{c^{4}+4a_{1}}}}{%
2}s\right) +  \notag \\
&&+v_{3}\exp \left( +i\frac{\sqrt{2c^{2}-2\sqrt{c^{4}+4a_{1}}}}{2}s\right)
+v_{4}\exp \left( -i\frac{\sqrt{2c^{2}-2\sqrt{c^{4}+4a_{1}}}}{2}s\right) .
\end{eqnarray}

\subsection{Scaling solutions for the forced Euler-Bernoulli equation.\newline}

We continue by presenting the reduction with the scaling symmetries for the
Euler-Bernoulli equation for the power-law and the exponential sources $%
f_{2}\left( u\right) $ and $f_{3}\left( u\right) $. For simplicity and
without loss of generality we select $b=0$.

For the power-law source $f_{2}\left( u\right) =au^{n}$, the application of
the Lie point symmetry $\Gamma _{3}^{f_{2}}$ provides the reduced
fourth-order ode,
\begin{equation}
4\alpha \beta v^{\prime \prime \prime \prime }+s^{2}v^{\prime \prime }+\frac{%
3n+5}{n-1}sv^{\prime }+\frac{8(1+n)v}{\left(n-1\right) ^{2}}-4av^{n}=0,
\end{equation}%
in which $s=\frac{x}{\sqrt{t}}$ and $u\left( t,x\right) =v\left( s\right) t^{%
\frac{2}{n-1}}$.

On the other hand, for the exponential source $f_{3}\left( u\right) =e^{au}$%
, the reduced equation given by the scaling symmetry is,
\begin{equation}
4\alpha \beta v^{\prime \prime \prime \prime }+s^{2}v^{\prime \prime
}+3sav^{\prime }+4ae^{av}+8=0,
\end{equation}%
where now $s=\frac{x}{\sqrt{t}}$ and $u\left( t,x\right) =-\frac{2}{a}\ln
\left( t\right) +v\left( s\right) $.

\section{Singularity analysis}

The third-order ode that we apply the singularity analsyis is (\ref{02i}) or
\begin{eqnarray}
24\nu xy(x)+y(x)^{2}+156\nu x^{2}y(x)^{2}+112\nu x^{3}y(x)^{3}+16\nu
x^{4}y(x)^{4}+y^{\prime }(x)+156\nu x^{2}y^{\prime}(x)\notag \\+336\nu x^{3}y(x)y^{\prime}(x)+
+96\nu x^{4}y(x)^{2}y^{\prime}(x)+48\nu x^{4}y^{\prime 2}(x)+112\nu x^{3}y^{\prime \prime}(x)\notag \\
+64\nu x^{4}y(x)y^{\prime \prime}(x)+16\nu x^{4}y^{\prime\prime\prime}(x)&=&0,\notag \\
\end{eqnarray}%
where $g\left( h\right) =y\left( x\right) $ ,~$h=x$ and $\alpha \beta =v$.
We apply the ARS algorithm\ \cite{ars1,ars2,ars3} and we make the usual
substitution to obtain the leading-order behaviour \cite{Article16},
\begin{equation}
y\rightarrow a(x-x_{0})^{p},
\end{equation}%
which provides

\begin{eqnarray}
32a\nu px^{4}(x-x_{0})^{-3+p}-48a\nu p^{2}x^{4}(x-x_{0})^{-3+p}+16a\nu
p^{3}x^{4}(x-x_{0})^{-3+p}-112a\nu px^{3}(x-x_{0})^{-2+p}  \notag \\
+112a\nu p^{2}x^{3}(x-x_{0})^{-2+p}+ap(x-x_{0})^{-1+p}+156a\nu
px^{2}(x-x_{0})^{-1+p}+24a\nu x(x-x_{0})^{p}  \notag \\
+a^{2}(x-x_{0})^{2p}+156a^{2}\nu x^{2}(x-x_{0})^{2p}+112a^{3}\nu
x^{3}(x-x_{0})^{3p}+16a^{4}\nu x^{4}(x-x_{0})^{4p}  \notag \\
-64a^{2}\nu px^{4}(x-x_{0})^{-2+2p}+112a^{2}\nu
p^{2}x^{4}(x-x_{0})^{-2+2p}+336a^{2}\nu px^{3}(x-x_{0})^{-1+2p}\notag \\
+96a^{3}\nu px^{4}(x-x_{0})^{-1+3p}&=&0.\notag \\
\end{eqnarray}%
From the latter it is evident that $p\rightarrow -1$. Hence,%
\begin{eqnarray}
-\frac{96a\nu x^{4}}{(x-x_{0})^{4}}+\frac{176a^{2}\nu x^{4}}{(x-x_{0})^{4}}%
-\frac{96a^{3}\nu x^{4}}{(x-x_{0})^{4}}+\frac{16a^{4}\nu x^{4}}{(x-x_{0})^{4}%
}+\frac{224a\nu x^{3}}{(x-x_{0})^{3}}  \notag \\
-\frac{336a^{2}\nu x^{3}}{(x-x_{0})^{3}}+\frac{112a^{3}\nu x^{3}}{%
(x-x_{0})^{3}}-\frac{a}{(x-x_{0})^{2}} \\
+\frac{a^{2}}{(x-x_{0})^{2}}-\frac{156a\nu x^{2}}{(x-x_{0})^{2}}+\frac{%
156a^{2}\nu x^{2}}{(x-x_{0})^{2}}+\frac{24a\nu x}{x-x_{0}}.  \notag
\end{eqnarray}%
We extract the obvious dominant terms,
\begin{equation}
-\frac{96a\nu x^{4}}{(x-x_{0})^{4}}+\frac{176a^{2}\nu x^{4}}{(x-x_{0})^{4}}-%
\frac{96a^{3}\nu x^{4}}{(x-x_{0})^{4}}+\frac{16a^{4}\nu x^{4}}{(x-x_{0})^{4}},
\end{equation}%
and solve for $a$,
\begin{equation}
(-3+a)(-2+a)(-1+a)a=0,
\end{equation}%
to obtain,
\begin{equation}
a\rightarrow 0~,a\rightarrow 1~,a\rightarrow 2~,a\rightarrow 3.
\end{equation}

In order to find the resonances, we substitute,%
\begin{equation}
y\rightarrow a(x-x_{0})^{-1}+m(x-x_{0})^{-1+s},
\end{equation}%
and we linearize around $m$.

Then the usual $x\rightarrow z+x_{0}$, simplify the calculations and provide
the dominant terms factor as,
\begin{equation}
16\nu (-3+2a+s)\left( 2-6a+2a^{2}-3s+2as+s^{2}\right) x_{0}^{4}z^{-4+s},
\end{equation}%
and for the three values of $a$, we obtain the three sets of resonances, for
each value of the coefficient term \ $a:$%
\begin{equation*}
a=1:s\rightarrow -1,~s\rightarrow 1,~s\rightarrow 2,
\end{equation*}
\begin{equation*}
~a=2:s\rightarrow -2,~s\rightarrow -1,~s\rightarrow 1,
\end{equation*}
and%
\begin{equation*}
a=3:s\rightarrow -3,~s\rightarrow -2,~s\rightarrow -1.
\end{equation*}

For $a=1$, the solution is expressed in terms of a Right Painlev\'{e} Series,
for $a=3$, in terms of a Left Painlev\'{e} Series and for $a=2$, in terms of a
Mixed Painlev\'{e} Series.

We commence the consistency test. For the Right Painlev\'{e} Series, we write
\begin{equation}
y\rightarrow
(x-x_{0})^{-1}+F_{0}+F_{1}(x-x_{0})+F_{2}(x-x_{0})^{2}+F_{3}(x-x_{0})^{3}+...
\end{equation}%
where $F_{0}$ and $F_{1}$ are the second and third constants of integration.
The output is enormous and hence omitted.

The substitution $x\rightarrow z+x_{0}$, just makes it easier to collect like
powers. The terms in $z^{-1}$ are,
\begin{eqnarray}
\frac{2F_{0}}{z}+\frac{24\nu x_{0}}{z}+\frac{312F_{0}\nu x_{0}^{2}}{z}+%
\frac{336F_{0}^{2}\nu x_{0}^{3}}{z}+\frac{336F_{1}\nu x_{0}^{3}}{z}+  \notag
\\
+\frac{64F_{0}^{3}\nu x_{0}^{4}}{z}+\frac{192F_{0}F_{1}\nu x_{0}^{4}}{z}+%
\frac{128F_{2}\nu x_{0}^{4}}{z}+......&=&0.
\end{eqnarray}%
This is solved to give,
\begin{eqnarray}
64\nu x_{0}^{4}F_{2} &\rightarrow &-F_{0}-12\nu x_{0}-156F_{0}\nu
x_{0}^{2}-168F_{0}^{2}\nu x_{0}^{3}+  \notag \\
&&-168F_{1}\nu x_{0}^{3}-32a_{0}^{3}\nu x_{0}^{4}-96F_{0}F_{1}\nu x_{0}^{4}.
\end{eqnarray}%
The expression for $F_{2}$ is substituted into the major output,

\begin{eqnarray}
-7F_{0}^{2}+3F_{1}-240\nu -\frac{22F_{0}}{x_{0}}-2880F_{0}\nu
x_{0}-3780F_{0}^{2}\nu x_{0}^{2}-2220F_{1}\nu x_{0}^{2}+  \notag \\
-1680F_{0}^{3}\nu x_{0}^{3}-1680F_{0}F_{1}\nu x_{0}^{3}-240F_{0}^{4}\nu
x_{0}^{4}+ \notag \\
-480F_{0}^{2}F_{1}\nu x_{0}^{4}+240F_{1}^{2}\nu x_{0}^{4}+480F_{3}\nu
x_{0}^{4}&=&0, \notag \\
\end{eqnarray}%
and the coefficient of the constant term is solved to give $F_{3}$ as,
\begin{eqnarray}
480\nu x_{0}^{5}F_{3} &=&22F_{0}+7F_{0}^{2}x_{0}-3F_{1}x_{0}+240\nu
x_{0}+2880F_{0}\nu x_{0}^{2}+3780F_{0}^{2}\nu x_{0}^{3}  \notag \\
&&+2220F_{1}\nu x_{0}^{3}+1680F_{0}^{3}\nu x_{0}^{4}+1680F_{0}F_{1}\nu
x_{0}^{4}+240F_{0}^{4}\nu x_{0}^{5}
+480F_{0}^{2}F_{1}\nu x_{0}^{5}-240F_{1}^{2}\nu x_{0}^{5}. \notag \\
\end{eqnarray}%
Thus there is not a problem with the determination of the coefficients of
the terms in the Right Painlev\'{e} Series. As the terms, not included in
those which are dominant are less dominant, there cannot be a Left Painlev%
\'{e} Series. The possibility of the existence of a Mixed Painlev\'{e}
Series is moot due to the practical difficulty of calculating coefficients.
Consequently, equation (\ref{02i}) is integrable according to the Painlev\'{e}
test

\section{Conservation laws}
The Ibragimov's theory of nonlinear self-adjointness details to construct conservation laws for a scalar pde\cite{bk37, bk39,bk38}. Our first step is to verify the self-adjointness condition on the various form of beam and later on compute the conservation laws. The preliminaries can be easily accessed from\cite{bk37}.\newline

The main motivation behind using the Ibragimov's approach is to obtain the conservation laws to deduce certain special solutions for the Beam equations following the methodology specified by  Cimpoiasu \cite{Acta04} where the author have used the nonlinear self-adjointness method to compute solutions for the Rossby waves. The Noether's theorem can be easily applied to obtain the conserved terms but it is our intuition that the non-local conserved terms as obtained using the Ibragimov's method can contribute in obtaining new solutions in a different subspace of the complex plane.
The main objective is to deduce new solutions using the point symmetries, singularities and conservation laws.
For instance, the Euler-Bernoulli equation does imply the existence of series type solutions through the method of Singularity analysis as mentioned in Section $5$ for equation (\ref{02i}).\\

Let the scalar PDE admit the following generators of the infinitesimal
transformation,
\begin{equation}
V=\xi ^{i}(x,u,u_{i},..)\frac{\partial }{\partial x^{i}}+\eta
(x,u,u_{i},...)\frac{\partial }{\partial u}.
\end{equation}%
Then the scalar PDE and its adjoint equation, as defined above, admits the
conservation law
\begin{eqnarray*}
C^{i}=\xi ^{i}\mathit{L}+W\left(\frac{\partial {\mathit{L}}}{\partial _{u_{i}}}%
-D_{j}\left(\frac{\partial {\mathit{L}}}{\partial _{u_{ij}}}\right)+D_{j}D_{k}\left(\frac{%
\partial {\mathit{L}}}{\partial _{u_{ijk}}}\right)-...\right)+D_{j}(W)\left(\frac{\partial {%
\mathit{L}}}{\partial _{u_{ij}}}-D_{k}\left(\frac{\partial {\mathit{L}}}{\partial
_{u_{ij}}}\right)+...\right)\notag\\+D_{j}D_{k}(W)\left(\frac{\partial {\mathit{L}}}{\partial
_{u_{ijk}}}-...\right),
\end{eqnarray*}%
where $W=\eta -\xi ^{i}u_{i}$ and $\mathit{L}$ denotes the Lagarangian of the corresponding form of the beam equation.
For the Euler-Bernoulli, Rayleigh and Timshenko-Prescott forms the Lagarangians are as follows
\begin{eqnarray*}
\mathit{L}&=&q(t,x)(u_{tt}+\alpha \beta u_{xxxx}),\nonumber\\
\mathit{L}&=&q(t,x)(\alpha \beta u_{xxxx} + u_{tt} - \beta u_{xxtt}),\nonumber\\
\mathit{L}&=&q(t,x)(\alpha \beta u_{xxxx} +u_{tt} - \beta (1 + \epsilon)u_{xxtt}+\frac{\epsilon \beta u_{tttt}}{\alpha}),\nonumber\\
\end{eqnarray*}
where $q(t,x)$ is the new dependent variable. To verify the non-linear self adjointness the substitution of $q(t,x)=\phi(t,x,u)$, to the adjoint equation of (\ref{01a}), (\ref{01c}) and (\ref{01e}) must satisfy for all solutions $u$ of those equations.  The possible values of $\phi(t,x,u)$ is a constant term, let say, $A_{0}$ and $A_{1} u(t,x)+A_{2}$, where $A_{1}$ and $A_{2}$ are arbitrary constants. A complete description of this method can be obtained from \cite{bk37}.

\subsection{Conservation laws for the various form of beam.\\}
 With respect to, each of the symmetry in (\ref{01a}), we compute the nonzero conservation laws.\newline
 For $\Gamma_{1a}$ the conservation components are
 \begin{align}
 c^{t}&=\begin{aligned}[t]&u_{x}\phi_{t}-u_{xt}\phi(t,x,u),\nonumber\\\end{aligned}\\
 c^{x}&=\begin{aligned}[t]&\phi(t,x,u)(u_{tt}+\alpha\beta u_{xxxx})+\alpha\beta u_{x}\phi_{xxxx}-\alpha\beta u_{xx}\phi_{xx}
 +\alpha\beta \phi_{x}u_{xxx}-u_{xxxx}\phi(t,x,u).\nonumber\\\end{aligned}\\
 \end{align}
 For $\Gamma_{2a}$,
 \begin{align}
 c^{t}&=\begin{aligned}[t]&\phi(t,x,u)(u_{tt}+\alpha\beta u_{xxxx})+u_{t}\phi_{t}-u_{tt}\phi(t,x,u),\nonumber\\\end{aligned}\\
 c^{x}&=\begin{aligned}[t]&\alpha\beta(u_{t}\phi_{xxx}-u_{xt}\phi_{xx}-u_{xxt}\phi_{x}-u_{xxxt}\phi(t,x,u)).\nonumber\\\end{aligned}\\
 \end{align}
 For  $\Gamma_{3a}$,
 \begin{align}
 c^{t}&=\begin{aligned}[t]&-u\phi_{t}+u_{t}\phi(t,x,u),\nonumber\\\end{aligned}\\
 c^{x}&=\begin{aligned}[t]&\alpha\beta(u_{x}\phi_{xx}-u\phi_{xxx}+u_{xx}\phi_{x}+u_{xxxx}).\nonumber\\\end{aligned}\\
 \end{align}
 For $\Gamma_{4a}$,
 \begin{align}
 c^{t}&=\begin{aligned}[t] &2t(\phi(t,x,u)(u_{tt}+\alpha\beta u_{xxxx}))+\phi_{t}(2tu_{t}+xu_{x})
 -\phi(t,x,u)(2tu_{tt}+2u_{t}+xu_{xt}),\nonumber\\\end{aligned}\\
 c^{x}&=\begin{aligned}[t] &x(\phi(t,x,u)(u_{tt}+\alpha\beta u_{xxxx}))+\alpha\beta \phi_{xxx}(2tu_{t}+xu_{x})
 -\alpha\beta \phi_{xx}(2tu_{xt}+xu_{xx}+u_{x})\nonumber\\
 &+\alpha\beta \phi_{x}(2tu_{xxt}+xu_{xxx}+2u_{xx})
 -\alpha\beta\phi(t,x,u)(2tu_{xxxt}+xu_{xxxx}+3u_{xxx}).\nonumber\\\end{aligned}\\
 \end{align}
F0r $\Gamma_{5a}$,
\begin{align}
c^{t}&=\begin{aligned}[t]&-a(t,x)\phi_{t}+a_{t}\phi(t,x,u),\nonumber\\\end{aligned}\\
c^{x}&=\begin{aligned}[t]&-\alpha\beta a(t,x)\phi_{xxx}+\alpha\beta a_{x}\phi_{xx}-\alpha\beta a_{xx}\phi_{x}
+\phi(t,x,u)a_{xxx}.\nonumber\\\end{aligned}\\
\end{align}

Next we compute the conservation laws of equation (\ref{01c}), with respect to its symmetries.\newline
$\Gamma_{1b}$ leads to the following conserved components,
\begin{align}
c^{t}&=\begin{aligned}[t]&\phi(t,x,u)(\alpha\beta u_{xxxx}+u_{tt}-\beta u_{xxtt})+u_{t}\phi_{t}-\phi(t,x,u)u_{tt},\nonumber\\\end{aligned}\\
c^{x}&=\begin{aligned}[t]&\alpha\beta(u_{t}\phi_{xxx}-u_{xt}\phi_{xx}+u_{xxt}\phi_{x}-u_{xxxt}\phi(t,x,u)).\nonumber\\\end{aligned}\\
\end{align}

For $\Gamma_{2b}$,
\begin{align}
c^{t}&=\begin{aligned}[t]&u_{x}\phi_{t}-u_{xt}\phi(t,x,u),\nonumber\\\end{aligned}\\
c^{x}&=\begin{aligned}[t]&\phi(t,x,u)(\alpha\beta u_{xxxx}+u_{tt}-\beta u_{xxtt})+\alpha\beta(u_{x}\phi_{xxx}-u_{xx}\phi_{xx}+u_{xxx}\phi_{x}
-\phi(t,x,u)u_{xxxx}).\nonumber\\\end{aligned}\\
\end{align}

For $\Gamma_{3b}$,
\begin{align}
c^{t}&=\begin{aligned}[t]&u_{t}\phi(t,x,u)-u\phi_{t},\nonumber\\\end{aligned}\\
c^{x}&=\begin{aligned}[t]&\alpha\beta(u_{x}\phi_{xx}-u\phi_{xxx}-u_{xx}\phi_{x}+u_{xxx}\phi(t,x,u)).\nonumber\\\end{aligned}\\
\end{align}

For $\Gamma_{4b}$,
\begin{align}
c^{t}&=\begin{aligned}[t]&b_{t}\phi(t,x,u)-b(t,x)\phi_{t},\nonumber
\end{aligned} \\
c^{x}&=\begin{aligned}[t]&\alpha\beta(b_{x}\phi_{xx}-b(t,x)\phi_{xxx}-b_{xx}%
\phi_{x}+b_{xxx}\phi(t,x,u)).\nonumber\end{aligned} \\
\end{align}

For the Timoshenko-Prescott form of the beam  (\ref{01e}), the conserved components are as follows:\newline
For $\Gamma_{1c}$,
\begin{align}
c^{t}&=\begin{aligned}[t]&\phi(t,x,u)\left(\alpha \beta u_{xxxx} +u_{tt} - \beta (1 + \epsilon)u_{xxtt}+\frac{\epsilon \beta u_{tttt}}{\alpha}\right)
+u_{t}\left(\phi_{t}+\frac{\epsilon\beta\phi_{ttt}}{\alpha}\right)-u_{tt}\left(\phi(t,x,u)+\frac{\epsilon\beta\phi_{tt}}{\alpha}\right)\nonumber\\
&+u_{ttt}\frac{\epsilon\beta\phi_{t}}{\alpha}-u_{tttt}\frac{\epsilon\beta\phi(t,x,u)}{\alpha},\nonumber\\ \end{aligned}\\
c^{x}&=\begin{aligned}[t]&\alpha\beta(u_{t}\phi_{xxx}-u_{xt}\phi_{xx}+u_{xxt}\phi_{x}-u_{xxxt}\phi(t,x,u)).\nonumber\\ \end{aligned}\\
\end{align}

For $\Gamma_{2c}$,
\begin{align}
c^{t}&=\begin{aligned}[t]&u_{x}\left(\phi_{t}+\frac{\phi_{ttt}\epsilon\beta}{\alpha}\right)-u_{xt}\left(\phi(t,x,u)+\frac{\phi_{tt}\epsilon\beta}{\alpha}\right)
+u_{xtt}\frac{\phi_{t}\epsilon\beta}{\alpha}-u_{xttt}\frac{\epsilon\beta}{\alpha},\nonumber\\ \end{aligned}\\
c^{x}&=\begin{aligned}[t]&\phi(t,x,u)\left(\alpha \beta u_{xxxx} +u_{tt} - \beta (1 + \epsilon)u_{xxtt}+\frac{\epsilon \beta u_{tttt}}{\alpha}\right)\nonumber\\
&+\alpha\beta(u_{x}\phi_{xxx}-u_{xx}\phi_{xx}+u_{xxx}\phi_{x}-\phi(t,x,u)u_{xxxx}).\nonumber\\ \end{aligned}\\
\end{align}

For $\Gamma_{3c}$,
\begin{align}
c^{t}&=\begin{aligned}[t]&-u\left(\phi_{t}+\frac{\epsilon\beta\phi_{ttt}}{\alpha}\right)+u_{t}\left(\phi(t,x,u)+\frac{\epsilon\beta\phi_{tt}}{\alpha}\right)
-u_{tt}\frac{\epsilon\beta\phi_{t}}{\alpha}+u_{ttt}\frac{\epsilon\beta\phi(t,x,u)}{\alpha},\nonumber\\ \end{aligned}\\
c^{x}&=\begin{aligned}[t]&\alpha\beta(u_{x}\phi_{xx}-u\phi_{xxx}-u_{xx}\phi_{x}+u_{xxx}\phi(t,x,u)).\nonumber\\ \end{aligned}\\
\end{align}

For $\Gamma_{4c}$,
\begin{align}
c^{t}&=\begin{aligned}[t]&-c(t,x)\left(\phi_{t}+\frac{\epsilon\beta\phi_{ttt}}{%
\alpha}\right)+c_{t}\left(\phi(t,x,u)+\frac{\epsilon\beta\phi_{tt}}{\alpha}\right)
-c_{tt}\frac{\epsilon\beta\phi_{t}}{\alpha}+c_{ttt}\frac{\epsilon\beta%
\phi(t,x,u)}{\alpha},\nonumber\end{aligned} \\
c^{x}&=\begin{aligned}[t]&\alpha\beta(c_{x}\phi_{xx}-c(t,x)\phi_{xxx}-c_{xx}%
\phi_{x}+c_{xxx}\phi(t,x,u)).\nonumber \end{aligned} \\
\end{align}

\section{Conclusion}

In this work, we focused on the algebraic properties for three different
forms of the beam equations with or without source. For the source free
equations we found that the Euler-Bernoulli equation is invariant under the
Lie algebra $(A_{3,3}\oplus A_{1})\oplus _{s}\infty A_{1},$ while, the
Rayleigh and Timoshenko-Prescott equations are invariant under the Lie
algebra $A_{3}\oplus _{s}\infty A_{1}$.

In the case of a isotropic source $f\left( u\right) $ we found that the
Euler-Bernoulli, Rayleigh and Timoshenko-Prescott equations are invariant
under the Lie algebra $2A_{1}$ for arbitrary source $f\left( u\right) $.
Moreover, for the Euler-Bernoulli beam equations the admitted Lie algebras
are $A_{3}\oplus _{s}\infty A_{1}$ for Linear $f\left( u\right) =au+b,~$\ $%
2A_{1}\oplus _{s}A_{1}$ for exponential or power-law functional form. For
the other two beam equations there are not specific functional forms of $%
f\left( u\right) $, where the equations admit different algebras. Therefore,
for the source-free equations we derived the conservation laws by applying
the Ibragimov's method.

We applied the Lie point symmetries to reduce the pdes and we proved that
the three beam equations provide exactly the same travelling-wave solutions.
The most important result of our paper is the reduction of Euler-Bernoulli
equation to a second-order equation, of the form of perturbed Painlev\'e-Ince
equation and to a third-order equation, which was studied by Chazy, Bureau
and Cosgrove. One of our subsequent paper will be on the singularity
analysis of the perturbed Painlev\'e-Ince equation. Moreover, our future work
also includes deriving further solutions of the different forms of beam
using conservation laws.

\subsection*{\textbf{Acknowledgement}}

AKH expresses grateful thanks to UGC (India) NFSC, Award No.
F1-17.1/201718/RGNF-2017-18-SC-ORI-39488 for financial support and Late Prof. K.M.Tamizhmani for the discussions which AKH had with him which formed the basis of this work.
PGLL acknowledges the support of the National Research Foundation of South
Africa, the University of KwaZulu-Natal and the Durban University of
Technology and thanks the Department of Mathematics, Pondicherry University,
for gracious hospitality.

\end{document}